\documentclass{article}

\pdfoutput=1

\usepackage{url}
\usepackage{graphicx}

\title{Social Dynamics of Science}

\author{Xiaoling Sun,$^{1,2}$ Jasleen Kaur,$^{2}$ Sta\v{s}a Milojevi\'{c},$^3$ \\
Alessandro Flammini,$^2$ Filippo Menczer$^2$
\\
$^1$ Department of Computer Science and Technology\\
Dalian University of Technology, China\\
$^2$ Center for Complex Networks and Systems Research\\
School of Informatics and Computing\\
$^3$ School of Library and Information Science\\
Indiana University, Bloomington, USA
}

\date{}

\begin{document}

\maketitle

\begin{abstract}
The birth and decline of disciplines are critical to science and society. However, no quantitative model to date allows us to validate competing theories of whether the emergence of scientific disciplines drives or follows the formation of social communities of scholars. Here we propose an agent-based model based on a \emph{social dynamics of science,} in which the evolution of disciplines is guided mainly by the social interactions among scientists. We find that such a social theory can account for a number of stylized facts about the relationships between disciplines, authors, and publications. 
These results provide strong quantitative support for the key role of social interactions in shaping the dynamics of science. A ``science of science'' must gauge the role of exogenous events, such as scientific discoveries and technological advances, against this purely social baseline.
\end{abstract}

\section{Introduction}
\label{intro}

Understanding the dynamics of science as a human endeavor --- the birth, evolution, and decline of disciplines --- is of critical importance for allocating resources and planning toward positive societal impact. For example, the emergence of new fields such as bioinformatics, nanophysics, information technology, quantum computing, and data science promises ``converging technologies'' with unparalleled potential to influence our lives. This paper is about modeling the dynamic evolution of scientific disciplines. 

Efforts to describe, explain and predict different aspects of science have intensified in recent years~\cite{BornerS09, BornerGSB11, scharnhorst2012models} spanning a wide range of theoretical, mathematical, statistical and computational approaches. A number of models of the dynamics of science have explored the continually changing disciplinary relations. Whether the new disciplines or fields are the results of branching of the old ones due to growth and new discoveries~\cite{price1986solla,mulkay1975three}, or go through a ``specialization-fragmentation-hybridization" cycle~\cite{dogan1990creative}, or grow through the synthesis of elements of pre-existing ones~\cite{etzkowitz2000dynamics}, these models point to the self-organizing development of science exhibiting growth and emergent behavior~\cite{noyons1998monitoring, vanraan1990fractal, van2000growth}. 

Kuhn's \emph{cognitive} theory emphasizes the role of observations not explained by previous paradigms~\cite{kuhn1996structure}. Other scholars~\cite{crane1972invisible, wagner2008new} emphasize the formation of \emph{social groups} of scientists as the driving force behind the evolution of disciplines. These models, however, are difficult to validate empirically or lack explanations of the processes leading to the empirical patterns they describe. 

Agent-based models~\cite{payette2012agent} allow one to generate macroscopic predictions from micro-level mechanisms guiding the behavior of individuals, thus providing a powerful approach to model the emergence of disciplines. While this approach has been used to create models of science dynamics~\cite{gilbert1997simulation, Borner:2004kx, Watts2011}, the focus was primarily on coauthorship, publication, and citation behavior rather than the emergence of disciplines. 

Quantitative work on modeling the emergence of disciplines is lacking to date, owing in part to the difficulty of formally defining the notion of scientific field, and the consequent sparsity of data to inform and validate models.
Here we offer a first quantitative baseline model to explore the consequences of assuming a purely social basis of science dynamics, without explicit references to exogenous events such as scientific discoveries.  In our model agents represent scholars who choose their collaborators, while groups of  collaborating scholars represent scientific disciplines~\cite{palla2007quantifying}. The key idea behind our model is that new  scientific fields emerge from splitting and merging of these social communities. 
Our model thus defines a \emph{social dynamics of science,} in which the birth and evolution of disciplines is guided mainly by the social interactions among scientists. We find that such a social theory can account for several stylized facts about the relationships between disciplines, authors, and publications. 

\section{Model Description}
\label{modeling}

\begin{figure}[h]
\centering
\includegraphics[width=\textwidth]{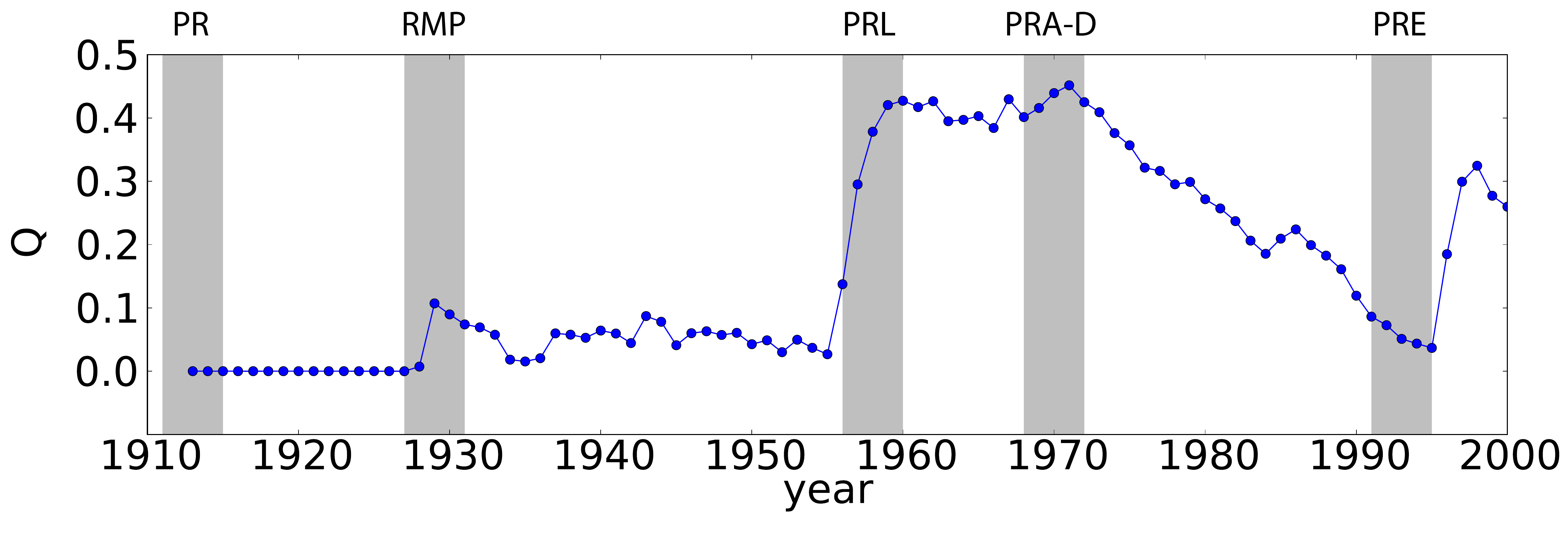}
\caption{Modularity $Q$ of APS journal-induced scholar communities. For each year $t$, we build a coauthor network based on the papers published in the 5-year time interval between $t-2$ and $t+2$. Such a network snapshot consists only of active authors, who published at least one paper in that time window. If an author published papers in more than one journal, we select the first journal in that period. The grey areas correspond to the introduction of major new journals.}
\label{fig:aps}
\end{figure}

The critical assumption of our model is the correspondence between the social dynamics of scholar communities and the evolution of scientific disciplines. To illustrate this intuition, let us look at the coauthorship network for papers published by the American Physical Society (APS). 
Using journals as proxies for scholarly communities, we can track the changes in community structure over time. 
Fig.~\ref{fig:aps} plots 
the modularity~\cite{Newman:2006fj} of the partition induced by the journals; higher values indicate a more clustered structure (see Methods). 

We observe noticeable changes in modularity around the introduction of new journals. Some of these changes suggest a scenario in which a new field emerges (e.g., quantum mechanics in the late 1920's), and a new journal captures the corresponding scholar community, leading to an increase in modularity. Interdisciplinary interactions across established areas lead to a decrease in modularity (e.g., prior to the introduction of \textit{Physical Review E} in the 1990's). These observations motivate a community detection approach to model the evolution of disciplines.

\begin{figure}
\centering
\includegraphics[width=\textwidth]{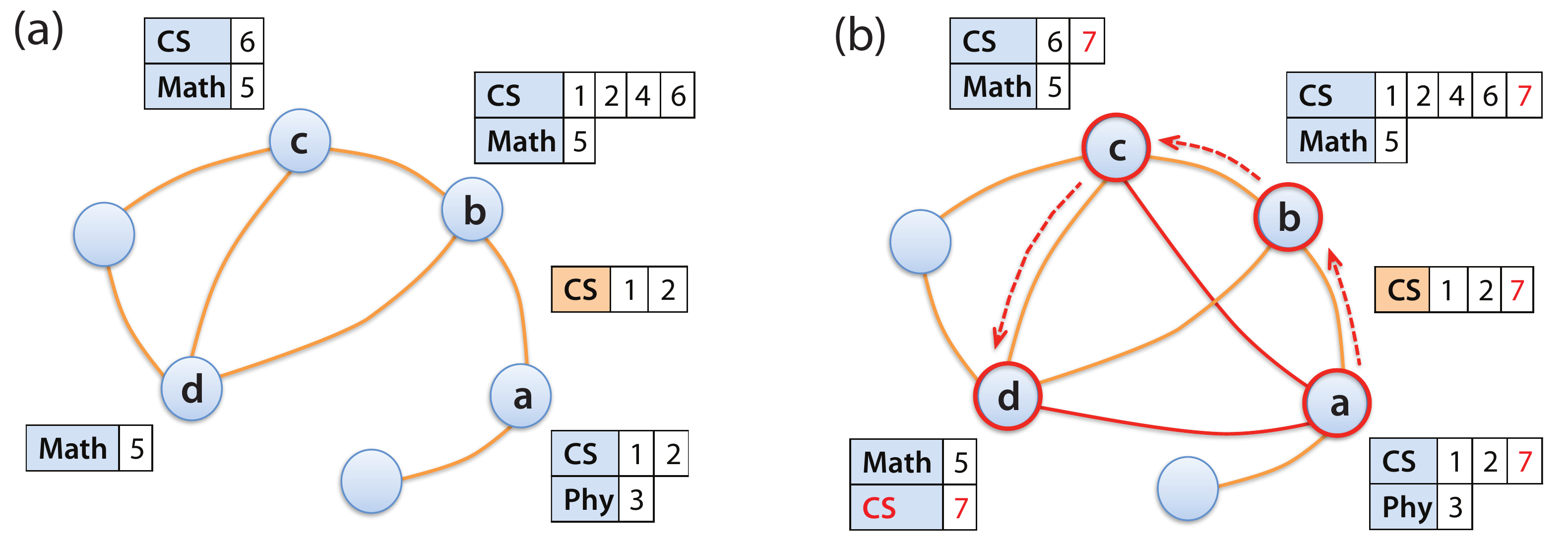}
\caption{(a) Illustration of the social network structure. Nodes and edges represent authors and their collaborations. They are annotated with lists of (co)authored papers grouped by scientific fields. For example, author $b$ has five papers including four in computer science (CS) and one in Math. Papers 1 and 2 are coauthored with $a$, papers 5 and 6 with $c$, and paper 5 with $d$. Paper 4 is authored by $b$ alone.
(b) Illustration of the random walk mechanism to select authors. For the new paper 7, the first author $a$ is chosen randomly and then walks to $b$ and $c$, stopping at $d$. These four authors become connected to each other if they have not collaborated before; for example, new edges connect $a$ to $c$ and $d$. Paper 7 acquires topics CS, Math and Physics (Phy). The main (majority) field of the paper, CS, diffuses across the coauthors, including $d$ who joins this discipline as a result.}
\label{fig:rw}
\end{figure}

In the proposed model, which we call SDS (for Social Dynamics of Science), we build a social network of collaborations whose nodes are authors, linked by coauthored papers as illustrated in Fig.~\ref{fig:rw}(a). Each author is represented by a list of disciplines indicating the scientific fields they have been working on, and every discipline has a list of papers. Similarly, each link is represented by a list of disciplines with associated papers describing the collaborations between two authors. 

There are three elements in the SDS model: papers, authors, and disciplines. The social network starts with one author writing one paper in one discipline. The network then evolves as new authors join, new papers are written, and new disciplines emerge over time.

At every time step, a new paper is added to the network. 
Its first author is chosen uniformly at random, so every author has a chance to publish a paper. 
In modeling the choice of collaborators, we aim to capture a few basic intuitions:
(i) authors who have collaborated before are likely to do so again;
(ii) authors with common collaborators are likely to collaborate with each other; 
(iii) it is easier to choose collaborators with similar than dissimilar background; and
(iv) authors with many collaborations have higher probability to gain additional ones~\cite{MERTON:1968zr, barabasi1999emergence}.
We model these behaviors through a biased random walk~\cite{Lovasz:1993random}, illustrated in Fig.~\ref{fig:rw}(b). 
The length of the random walk determines the number of coauthors. At each step in the walk, the author visits a node $i$  (starting with itself) and decides to stop with probability $p_w$, or to search for additional authors with probability $1-p_w$. In the latter case a neighbor $j$ is selected as coauthor according to the transition probability:
\begin{equation}
P_{ij} 
       = \frac{w_{ij}}{\sum_{k} w_{ik}}
\label{eq:tp}
\end{equation}
where $w_{ij}$ is the weight of the edge connecting authors $i$ and $j$, that is, the number of papers that $i$ and $j$ have written together. 
Note that the 
walk may result in a single author.

We propose a simple mechanism to model knowledge diffusion through collaboration in SDS: when authors write a paper together, they all contribute their knowledge. Therefore, a paper inherits the union of the author disciplines as topics. However, the discipline that is shared by the majority of authors is selected as the main topic of the paper (say, the publication venue) and diffuses across all the authors. Through the collaboration, authors acquire knowledge of and membership in this area.

At every time step, with probability $p_n$, we add a new author to the network with the new paper. The parameter $p_n$ regulates the ratio of papers to authors. 
The new author is the first author of the new paper. To generate other collaborators, an existing author is first selected uniformly at random as the first coauthor. Then the random walk procedure is followed to pick additional collaborators. The new author acquires the main topic of the paper. 

\begin{figure}
\centering
\includegraphics[width=\textwidth]{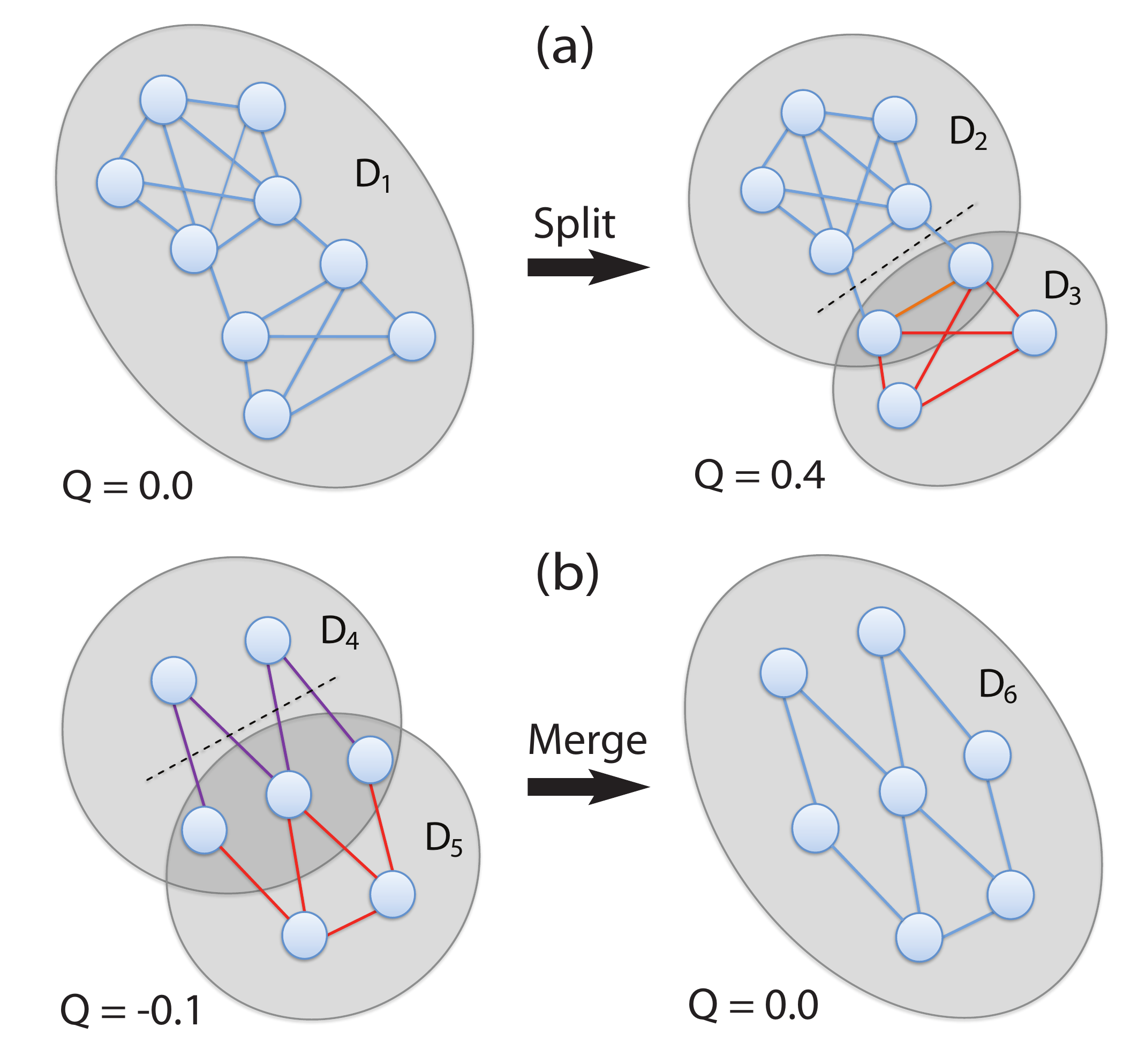}
\caption{Discipline evolution. (a) The coauthor network of discipline $D_1$ is split into two disciplines $D_2$ and $D_3$. The modularity increases from $Q = 0$ to $Q=0.4$. The dashed line indicates the partition of the network suggested by the community detection algorithm. Some nodes in the new discipline $D_3$ have also published papers with authors in $D_2$, and therefore belong to both disciplines.
(b) Two coauthor networks of disciplines $D_4$ and $D_5$ are merged into new discipline $D_6$. For authors in both original disciplines, we pick one based on the number of papers published in each discipline. The dashed line shows the resulting partition, with very low modularity $Q = -0.1$. The merged community $D_6$ has still low, but higher mudularity $Q=0$.}
\label{fig:evolve}
\end{figure}

We introduce a novel mechanism to model the evolution of disciplines by splitting and merging communities in the social collaboration network. The idea, motivated by the earlier observations from the APS data, is that the birth or decline of a discipline should correspond to an increase in the modularity of the network. Two such events may occur at each time step with probability $p_d$. 
The process is illustrated in Fig.~\ref{fig:evolve}.

For a \emph{split} event we select a random discipline with its coauthor network and decide whether a new discipline should emerge from a subset of this community. We partition the coauthor network into two clusters (see Methods). 
If the modularity of the partition is higher than that of the single discipline, there are more collaborations within each cluster than across the two. We then split the smaller community as a new discipline. 
In this case 
the papers whose authors are all in the new community are relabeled
to reflect the emergent discipline. 
Borderline papers with authors in both old and new disciplines are labeled according to the discipline of the majority of authors. Some authors may as a result belong to both old and new discipline. 

For a \emph{merge} event we randomly select two disciplines with at least one common author. If the modularity obtained by merging the two groups is higher than that of the partitioned groups, the collaborations across the two communities are stronger than those within each one. The two are then merged into a single new discipline. 
In this case 
all the papers in the two old disciplines are relabeled to reflect the new one. 

\section{Results}
\label{experiments}

To evaluate the predictive power of the SDS model we consider a number of \emph{stylized facts,} i.e., broad empirical observations that describe essential characteristics of the dynamic relationships between disciplines, scholars, and publications. Our model provides an explanation for the evolution of scientific fields if it can reproduce these empirical observations. The complex interactions of a changing group of scientists, their artifacts, and their disciplinary aggregations can be captured by the broad empirical distributions of six quantitative descriptors: 
the number of authors per paper $A_P$ (collaboration size);
the number of papers per author $P_A$ (scholar productivity);
the number of authors per discipline $A_D$ (discipline popularity);
the number of disciplines per author $D_A$ (scholar interdisciplinary effort);
the number of papers per discipline $P_D$ (discipline productivity); and
the number of disciplines per paper $D_P$ (publication breadth).

To validate the SDS model, one would ideally require a real-world dataset mapping the three-way relationships between scholars, publications, and disciplines. Unfortunately, to the best of our knowledge, no such dataset is publicly available. As an alternative, we adopt three large datasets that each map a binary projection of these relationships: \emph{NanoBank}~\cite{zucker2007nanobank} to validate the relationship between authors and papers, \emph{Scholarometer}~\cite{Scholarometer-PONE} to study the relationship between authors and disciplines, and \emph{Bibsonomy}~\cite{benz2010social} to analyze the relationship between papers and disciplinary topics. The datasets are described in the Methods section.  The parameters $p_n$, $p_w$, and $p_d$ of our model are tuned to fit the quantitative descriptors of each dataset (see Methods).  

\begin{figure}
\centering
\includegraphics[width=\textwidth]{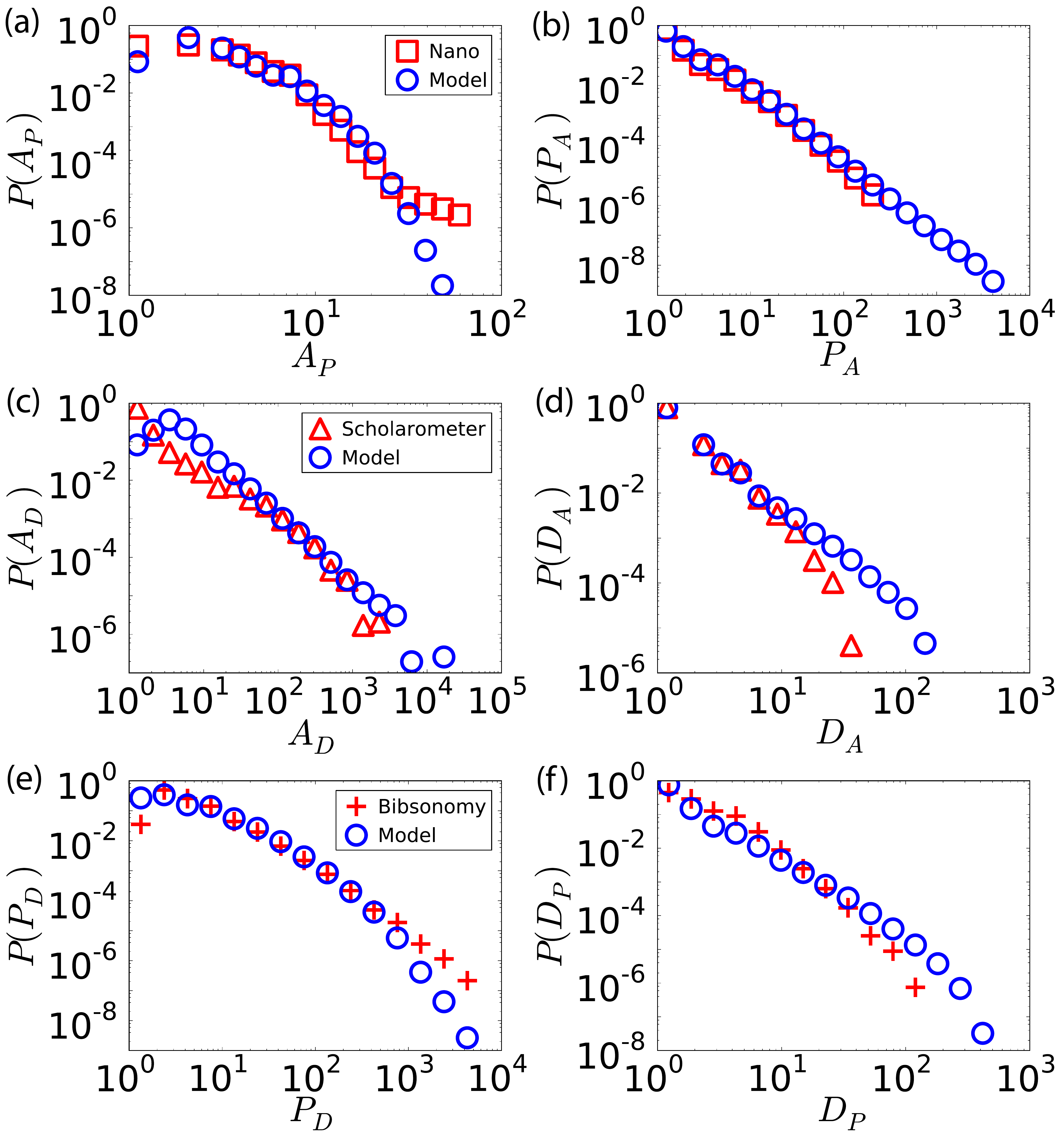}
\caption{Stylized facts characterizing relationships between authors, papers, and disciplines. We plot the distributions of (a)~authors per paper, (b)~papers per author, (c)~authors per discipline, (d)~disciplines per author, (e)~papers per discipline, and (f)~disciplines per paper. Blue circles represent the SDS predictions, while red symbols represent the empirical data from the three datasets. The results of the model are averaged over 10 runs.}
\label{fig:distributions}
\end{figure}

Fig.~\ref{fig:distributions} presents a close match between the real data and the predictions of our model. SDS reproduces the stylized facts about the relationships between scholars, publications, and disciplines, characterized by these six distributions. The exponential distribution of $A_P$ is captured by the random walk process. The broad distribution of scholar productivity $P_A$ is well accounted for by the bias in the random walk, which incorporates a kind of preferential attachment mechanism regulated by prior collaborations. The distributions of discipline popularity $A_D$ and productivity $P_D$ also display heavy tails, which cannot be attributed to a specific mechanism in the model; they emerge from the non-trivial interactions between (i) merging and splitting of the discipline communities and (ii) knowledge diffusion from the collaborations. The prediction is not as good for $D_A$: our model produces a relatively large number of highly interdisciplinary authors. One could correct this effect, for example, by requiring more than one paper in a discipline as a condition for membership. However, this would require an additional parameter and thus a more complicated model. Finally, The distribution of publication $D_P$ shows that there is a continuum in the breadth of papers, rather than a sharp separation between disciplinary and interdisciplinary work.

\begin{figure}
\centering
\includegraphics[width=\textwidth]{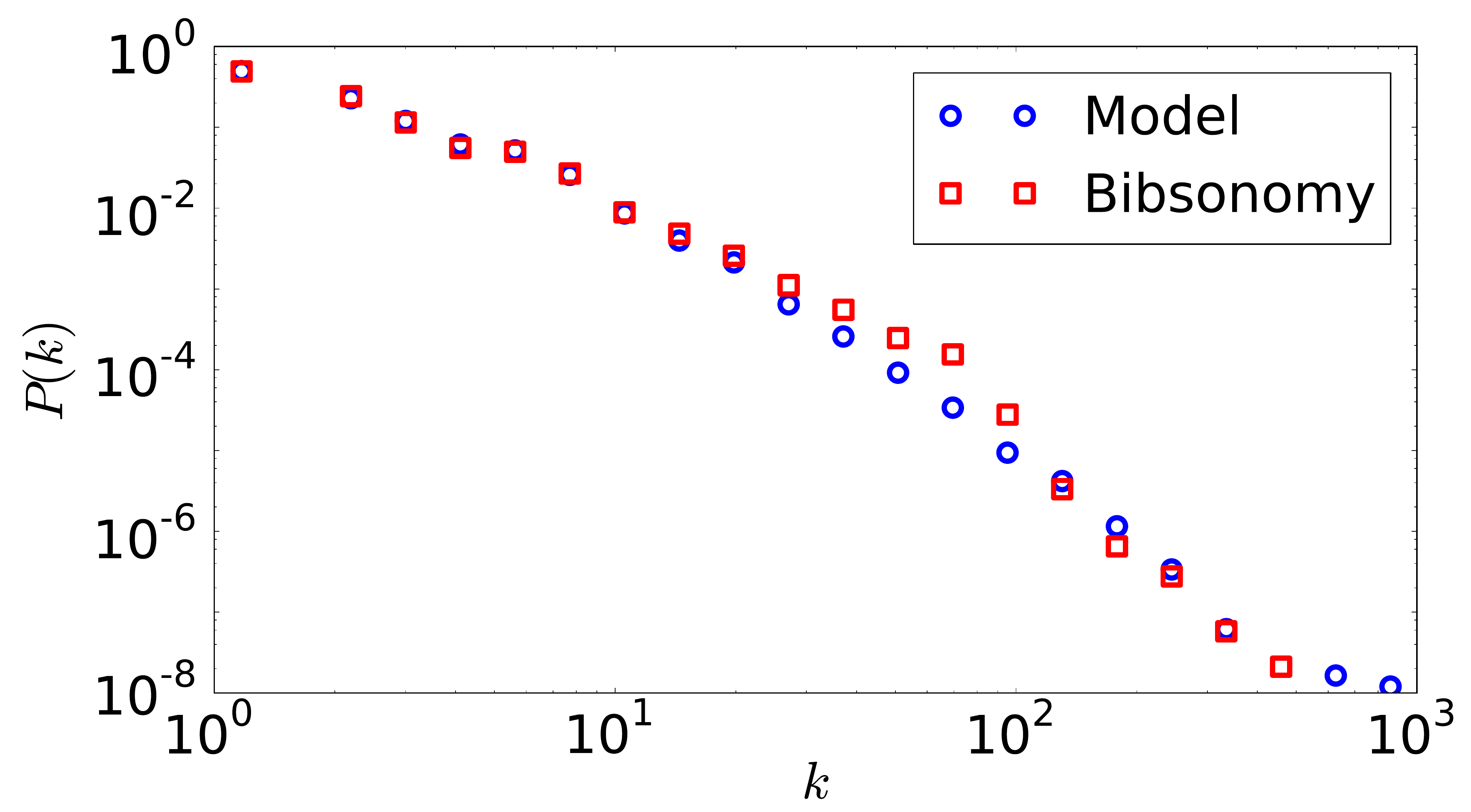}
\caption{Degree distribution of the coauthor network generated by the SDS model, compared to the empirical distribution from the Bibsonomy dataset. A similar match is also observed for other datasets (not shown). 
A few papers with more than 100 authors were excluded as they generate an anomaly in the tail; each such paper generates at least 100 nodes with degree at least 100.}
\label{fig:coauth-deg}
\end{figure}

These results focus on the relationships between disciplines, authors, and papers, for which there is little prior quantitative analysis. The coauthor network, on the other hand, has been studied extensively in the past~\cite{Newman:2001fk,Newman:2004qf}. As shown in Fig.~\ref{fig:coauth-deg}, the SDS model generates coauthor networks whose long-tailed degree distributions are consistent with the empirical data, as well as with those in the literature.

\section{Conclusions}
\label{conclusion}

We introduced an agent-based model to simulate the evolution of science as a process driven only by social dynamics. The model captures for the first time major stylized facts about the complex socio-cognitive interactions of a changing group of scholars, publications, and scientific communities. The SDS model is relatively simple when one considers the complexity of the science dynamics process being studied, yet powerful in its capability to reproduce the emergence of patterns similar to those observed in three real datasets about scientific production and fields. This provides us with strong quantitative support for the key role of social dynamics in shaping the birth, evolution, and decline of scientific disciplines.
Future ``science of science'' studies will have to gauge the role of scientific discoveries, technological advances, and other exogenous events in the emergence of new disciplines against this purely social baseline. 

\section{Methods}

Modularity~\cite{Newman:2006fj} measures the strength of a network partition into clusters of nodes. It compares the number of edges falling within groups with the expected number in an equivalent network from a null model with the same degree sequence but shuffled edges. Larger values indicate stronger community structure. 
Here we consider the weighted extension of modularity. Let $w_{ij}$ be the weight of an edge (number of coauthored papers) between nodes $i$ and $j$, and $W_{ij}$ its expected value. The weighted modularity is defined as: 
\begin{equation}
Q = \frac{1}{2m}\sum_{ij}[w_{ij} - W_{ij}] \delta(g_i,g_j)
\label{eq:modu}
\end{equation}
where $\delta(g_i,g_j) = 1$ if $g_i = g_j$ ($i$ and $j$ are in the same group) and 0 otherwise; $m$ is the sum of all edge weights in the network. $W_{ij} $ is computed as: 
\begin{equation}
W_{ij} =  \frac{s_i s_j}{2m}
\label{eq:expectp}
\end{equation}
where $s_i$ is the \emph{strength} or weighted degree of node $i$, $s_i = \sum_j w_{ij}$.  

When splitting disciplines, in practice, we use the leading eigenvector \linebreak method~\cite{Newman:2006qy} based on the (non-weighted) modularity matrix, as an efficient and effective algorithm to cluster a coauthor network into two groups. 

The APS dataset (Fig.~\ref{fig:aps}) was made available by the American Physical Society (\url{publish.aps.org/datasets/}). We consider the papers appearing in eight journals during the period of 1913-2000: \textit{Physical Review} (PR) 1913-1955, \textit{Review of Modern Physics} (RMP) 1929-2000, \textit{Physical Review Letters} (PRL) 1958-2000, \textit{Physical Review A, B, C, D} (PRA-D) 1970-2000, and \textit{Physical Review E} (PRE) 1993-2000.

The SDS model is validated against three datasets:

\begin{description}

\item[NanoBank]
(version Beta 1, released on May 2007)~\cite{zucker2007nanobank} is a digital library of bibliographic data on articles, patents and grants in the field of nanotechnology. A set of nanotechnology-related articles in NanoBank has been selected from the Science Citation Index Expanded, Social Sciences Citation Index, and Arts and Humanities Citation Index produced by the Institute for Scientific Information (now Thomson Reuters). Two document selection criteria have been used in the creation of NanoBank~\cite{zucker2007minerva}: (i) articles that contain some of the 379 terms identified by subject specialists as being ``nano-specific,'' and (ii) articles based on a probabilistic procedure for the automatic identification of terms. The database covers a 35-year period (1970-2004). This dataset is used to validate the relationship between authors and papers.

\item[Scholarometer]
(\url{scholarometer.indiana.edu}) is a social tool for scholarly services developed at Indiana University, with the goal of exploring the crowdsourcing approach for disciplinary annotations and cross-disciplinary impact metrics~\cite{hoang2010crowdsourcing,Scholarometer-PONE}. Users provide discipline annotations (tags) for queried authors, which in turn are used to compare author impact across disciplinary boundaries. The data collected by Scholarometer is available via an open API. We use this data to study the relationship between authors and disciplines.

\item[Bibsonomy]
(\url{www.bibsonomy.org}) is a system for sharing bookmarks and lists of literature~\cite{benz2010social}. Users annotate papers with tags. The dataset is free for research purposes. We downloaded a dump as of 2012-01-01 to analyze the relationship between papers and disciplines. To filter noise from many junk annotations, we removed the tags associated with fewer than 3 papers or more than 6,000 papers. 

\end{description}

The SDS model has three parameters: $p_n$ controls the number of papers per author; $p_w$ controls the number of authors per paper; finally, $p_d$ is the frequency of network split and merge events, and controls the number of disciplines.  
To generate predictions from numerical simulations of the model, we first tune these parameters to fit the properties of the three empirical  
datasets individually. 
Table~\ref{table:fitting} reports the main properties of the datasets and the matching model parameters. 
As shown in Table~\ref{table:basicstat}, the SDS model is capable of approximating the basic statistics of the empirical data.

\section*{Acknowlegments}
The work presented in this paper was performed while Xiaoling Sun was visiting the Center for Complex Networks and Systems Research (\url{cnets.indiana.edu}) at the Indiana University School of Informatics and Computing. Thanks to Diep Thi Hoang, Mohsen JafariAsbagh, and Lino Possamai for Scholarometer system development and helpful discussions, and John McCurley for editing assistance. We acknowledge support from Hongfei Lin at Dalian University of Technology, the China Scholarship Council, the Lilly Endowment, and NSF (award IIS-0811994) for funding the computing infrastructure that hosts the Scholarometer service.

\bibliography{Modeling}

\newpage

\section*{Figures and Tables}

\newpage

\begin{table}
\centering
\caption{Dataset properties and tuning of SDS model parameters. For each dataset we run the simulations until the empirical number of papers or authors is reached (shown in bold).}
\begin{tabular}{cccc}
\hline
					& NanoBank			& Scholarometer		& Bibsonomy	\\
\hline
Number of papers 	& $2.7 \times 10^5$		& $1.4 \times 10^6$	& $\mathbf{2.9 \times 10^5}$ \\
Number of authors	& $\mathbf{2.9 \times 10^5}$		& $\mathbf{2.2 \times 10^4}$	& $3.2 \times 10^5$ \\
Number of disciplines & n/a 				& $1.1 \times 10^3$	& $4.4 \times 10^4$ \\
$p_n$				& 0.90				& 0.04				& 0.80 \\
$p_w$ 				& 0.28				& 0.35				& 0.71 \\
$p_d$  				& n/a				& 0.01				& 0.50 \\
\hline	
\end{tabular}
\label{table:fitting}
\end{table}

\begin{table}
\centering
\caption{Basic statistics of empirical datasets compared with SDS model predictions. The reported values are averages, and standard deviations are obtained by 10 realizations of the model.}
\begin{tabular}{clcc}
\hline
Quantity		& Dataset		& Empirical		& SDS					\\ 
\hline
$A_P$		& NanoBank 		& 4.006			& $4.011 \pm 0.001$		\\
$P_A$		& NanoBank 		& 3.666			& $4.456 \pm 0.005$		\\	
$A_D$		& Scholarometer	& 45				& $60 \pm 10$		\\
$D_A$		& Scholarometer	& 2.2			& $3.5 \pm 0.4$			\\	
$P_D$		& Bibsonomy		& 24				& $22 \pm 1$			\\
$D_P$		& Bibsonomy		& 3.6			& $3.3 \pm 0.2$			\\	
\hline	
\end{tabular}
\label{table:basicstat}
\end{table}

\end{document}